\documentclass[twocolumn,pre,showpacs]{revtex4}
\usepackage{graphicx,color}
\usepackage{dcolumn}
\usepackage{amsmath,amssymb}
\def\FIGxW{2.88in}

\def\FIGxFIRST{
\begin{figure}[t]
 \mbox{\includegraphics[width=\FIGxW]{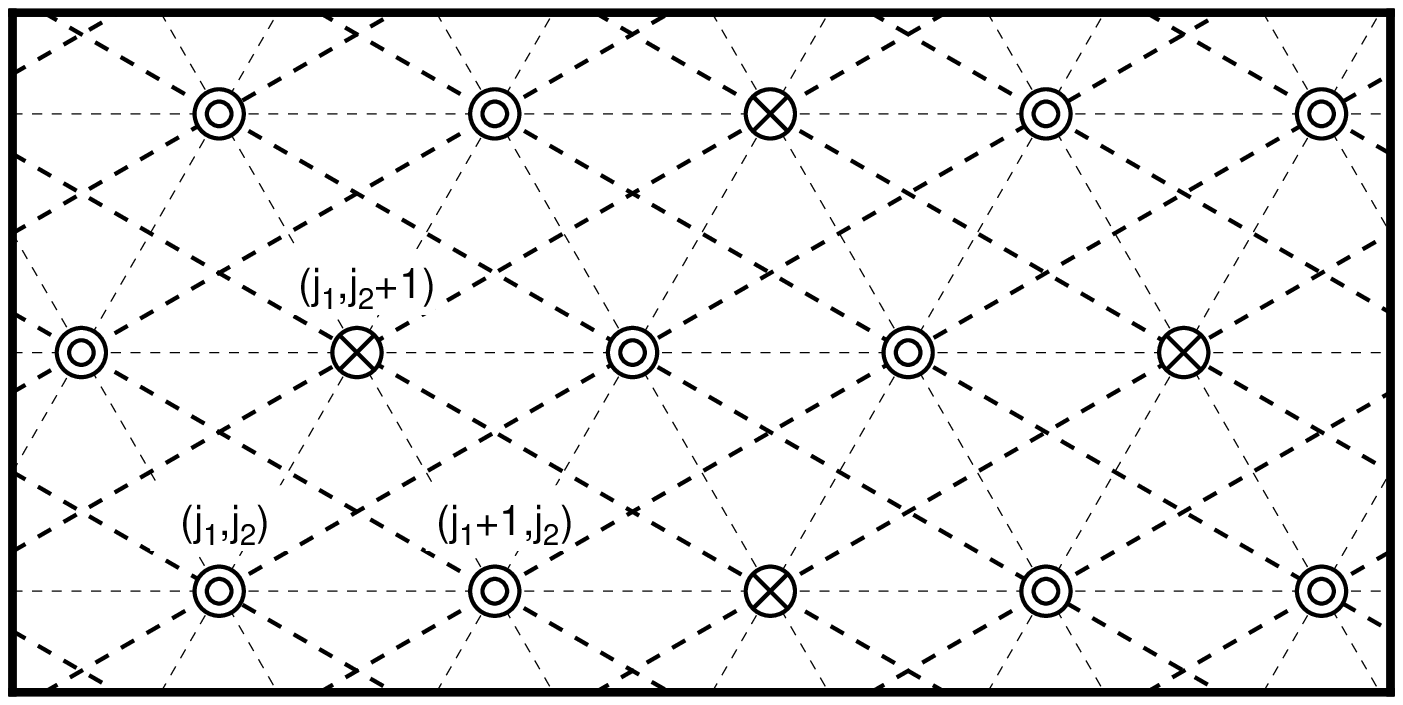}}
 \caption{
 The schematic representation of Hamiltonian\ (\ref{eq_Hamil}). 
 The $j$th site is specified by two integers $(j_1,j_2)$ as labeled in
 the figure.
 The dotted (dashed) lines show the NN AF (anisotropic NNN F) coupling. 
 An example of the spin configurations with the $\sqrt3\times\!\sqrt3$
 structure is exhibited, where the spins on two (one) of three sublattices
 $\Lambda_{0,1}$ ($\Lambda_{2}$) are up $\circledcirc$ (down
 $\otimes$).}
 \label{FIG1}
\end{figure} }

\def\FIGxSECOND{
\begin{figure}[t]
 \mbox{\includegraphics[width=\FIGxW]{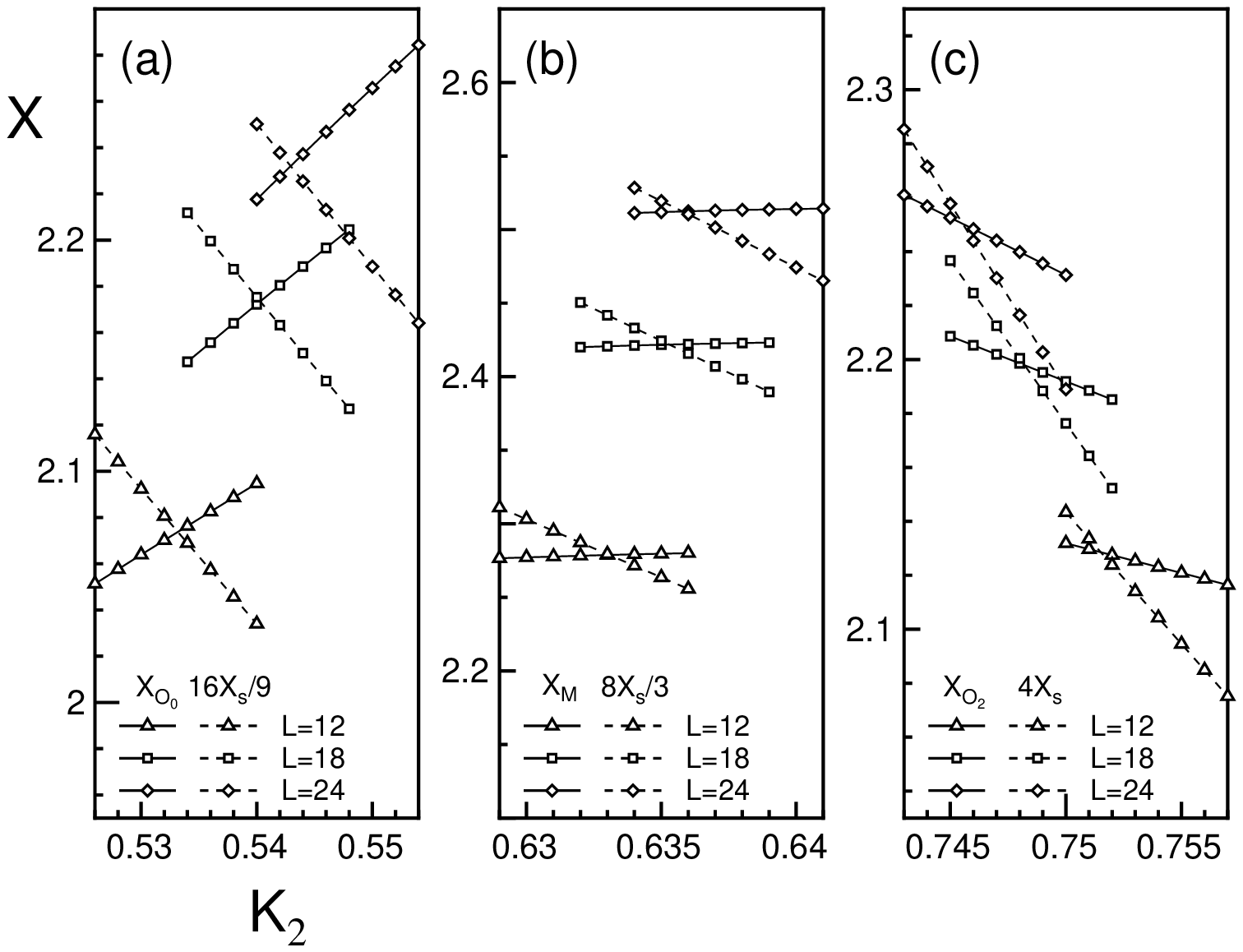}}
 \caption{
 The $K_2$ dependence of the scaled gaps at $K_1=1$ (the correspondences
 between marks and system sizes are given in these panels).
 (a) [(c)] shows $X_{O_0}$ and $16X_{s}/9$ ($X_{O_2}$ and $4 X_{s}  $). 
 Crossing points give finite-size estimates $v_{\rm U,L}(u,L)$.
 (b) gives data $X_{\cal M}$ and $8X_{s}/3$ whose crossing points
 estimate the self-dual line $v_{\rm sd}(u,L)$.}
 \label{FIG2}
\end{figure} }

\def\FIGxTHIRD{
\begin{figure}[t]
 \mbox{\includegraphics[width=\FIGxW]{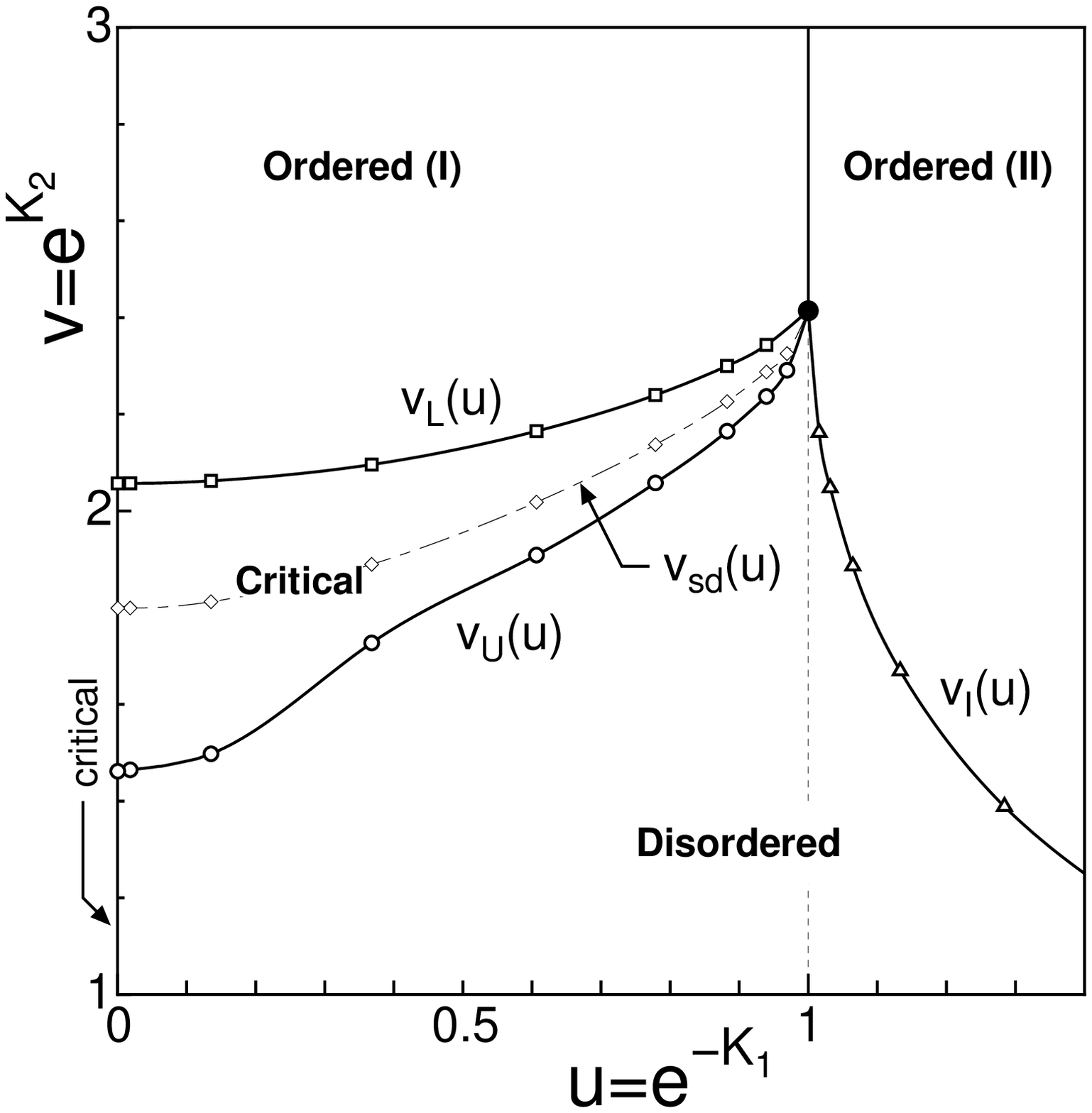}}
 \caption{
 The phase diagram.
 Open circles, squares and triangles with curves exhibit the
 BKT-transition lines $v_{\rm U}(u)$, $v_{\rm L}(u)$, and the second-order
 transition line $v_{\rm I}(u)$, respectively.
 The thick vertical line gives the first-order transition boundary.
 The filled circle at $(1,1+\sqrt{2})$ denotes the decoupling point with
 the three independent Ising criticality. 
 Diamonds with the dash-dotted line shows $v_{\rm sd}(u)$.}
 \label{FIG3}
\end{figure} }

\def\FIGxFOURTH{
\begin{figure}[t]
 \mbox{\includegraphics[width=\FIGxW]{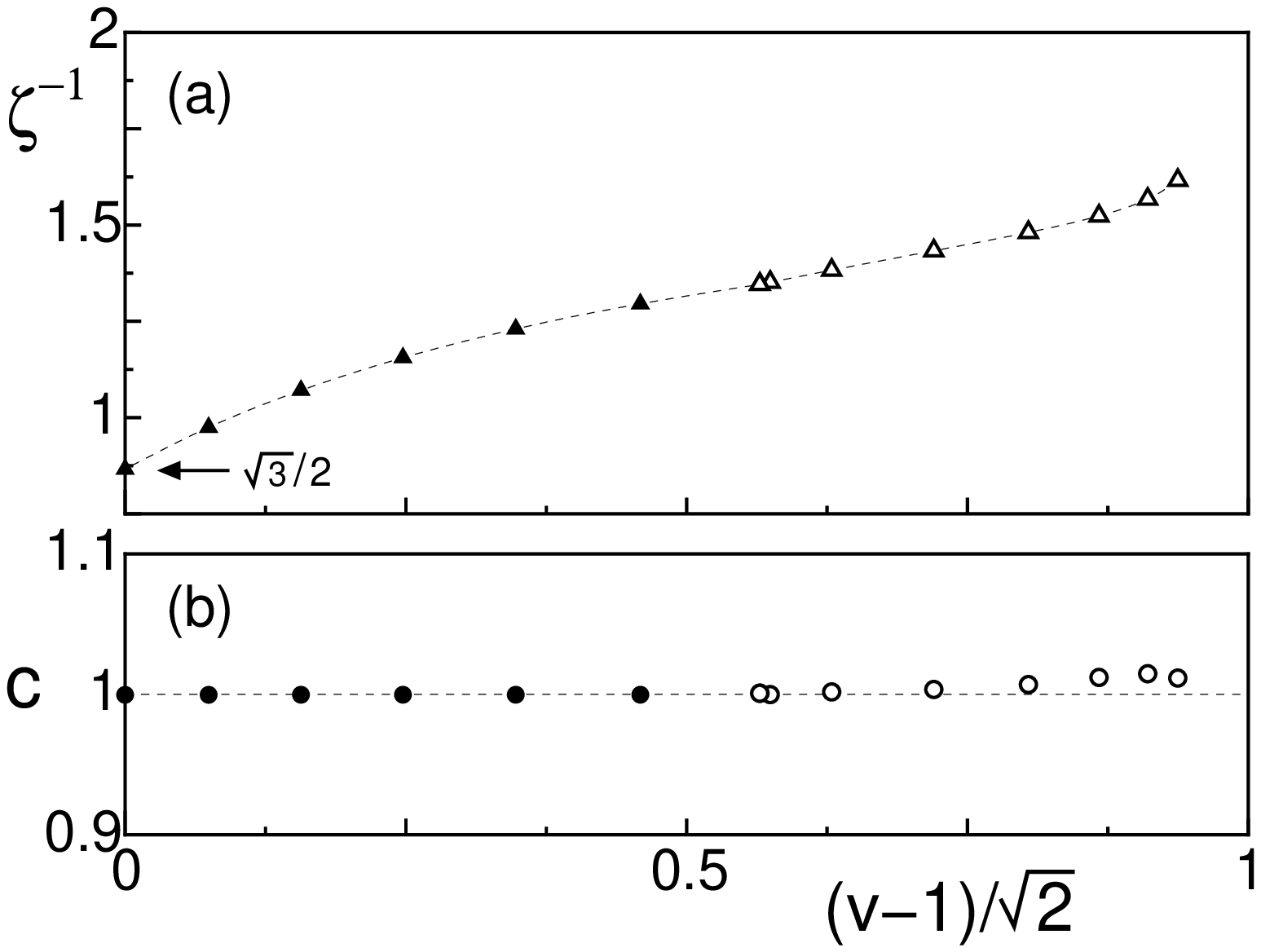}}
 \caption{
 The inverse effective geometric factor $\zeta^{-1}$ (the central charge
 $c$) versus $(v-1)/\sqrt2$ is plotted in the upper panel (a) [the lower
 panel (b)].
 Filled marks plot the data with $u=0$, and open ones along $v_{\rm
 sd}(u)$.
 $\zeta^{-1}$ takes the close value to $\sqrt3/2$ at $v=1$.}
 \label{FIG4}
\end{figure} }

\def\FIGxFIFTH{
\begin{figure}[t]
 \mbox{\includegraphics[width=\FIGxW]{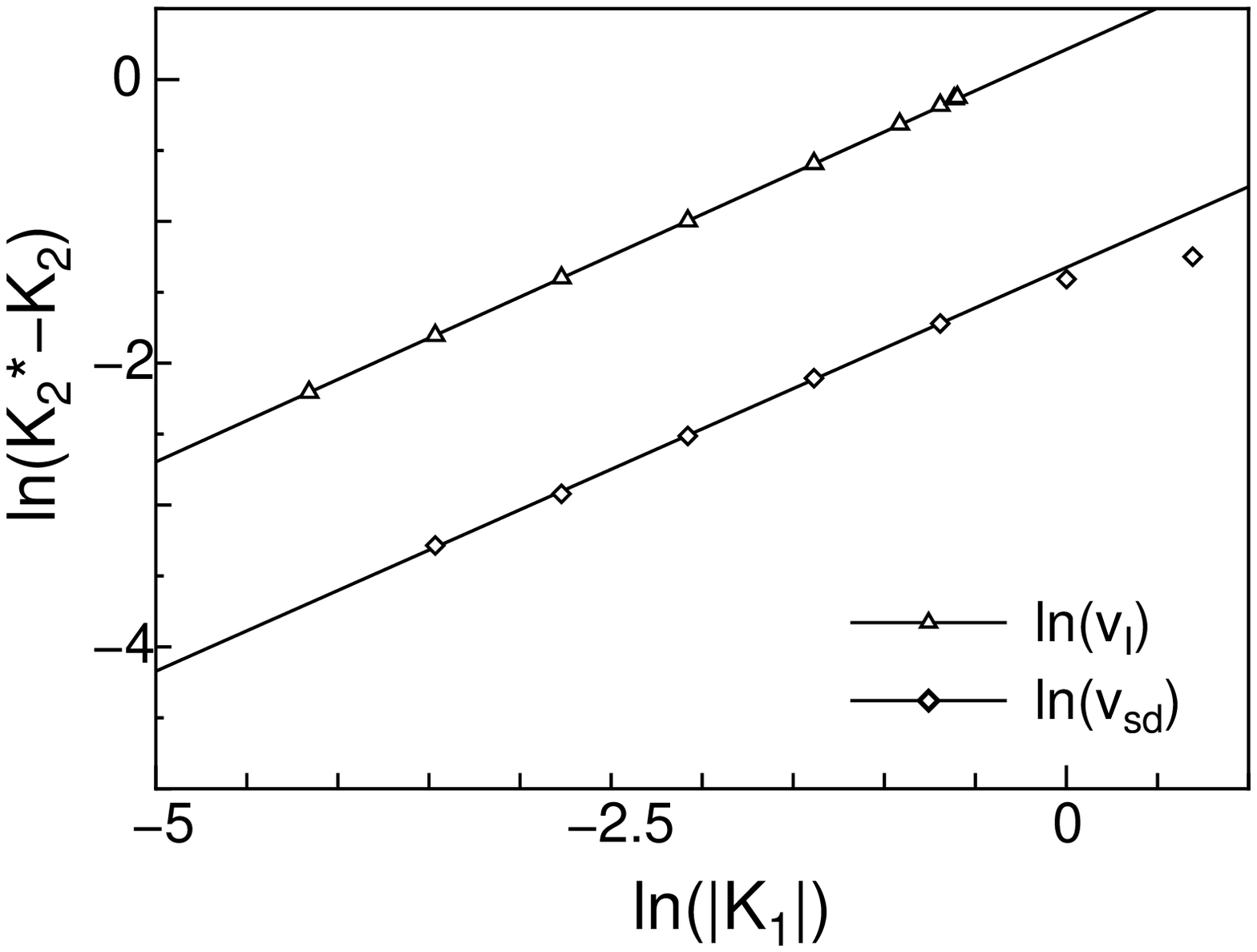}}
 \caption{
 Log-log plots of the phase boundary line $v_{\rm I}(u)$ and the self-dual
 line $v_{\rm sd}(u)$ around the critical decoupling point
 [we define $K^*_2=\ln(1+\sqrt2)$].
 The least-squares-fitting solid line for triangles (diamonds) estimates
 the crossover exponent
 $\phi_{\rm  F}\simeq 1.72$ ($\phi_{\rm AF}\simeq 1.76$)
 in the F (AF) case.}
 \label{FIG5}
\end{figure} }
\begin{document}
\title{
 Global phase diagram and six-state clock universality behavior in
 the triangular antiferromagnetic Ising model with anisotropic
 next-nearest-neighbor coupling:
 Level-spectroscopy approach
}
\author{
 Hiromi Otsuka
 and
 Yutaka Okabe
 }
\address{
 Department of Physics, Tokyo Metropolitan University, Tokyo 192-0397, Japan
 }
\author{
 Kiyohide Nomura
 }
\address{
 Department of Physics, Kyushu University, Fukuoka 812-8581, Japan
 }
\date{\today}
\begin{abstract}
 We investigate the triangular-lattice antiferromagnetic Ising model
 with a spatially anisotropic next-nearest-neighbor ferromagnetic
 coupling, which was first discussed by Kitatani and Oguchi.
 By employing the effective geometric factor, we analyze the scaling
 dimensions of the operators around the Berezinskii-Kosterlitz-Thouless
 (BKT) transition lines, and determine the global phase diagram.
 Our numerical data exhibit that two types of BKT-transition lines
 separate the intermediate critical region from the ordered and
 disordered phases, and they do not merge into a single curve in the
 antiferromagnetic region.
 We also estimate the central charge and perform some consistency checks
 among scaling dimensions in order to provide the evidence of the
 six-state clock universality. 
 Further, we provide an analysis of the shapes of boundaries based on
 the crossover argument.
\end{abstract}
\pacs{64.60.-i, 05.50.+q, 05.70.Jk}
\maketitle

\section{INTRODUCTION}\label{sec_INTROD}

 \FIGxFIRST

 The antiferromagnetic Ising model on the triangular lattice may be the
 simplest example to possess the frustration effects. 
 As its manifestation, the system does not complete the long-range order
 even in the ground state, but has the critical ground-state ensemble
 \cite{Wann50,Hout50,Husi50,Step70,Blot82}.
 This emerging criticality due to the frustration is the Gaussian type,
 where the Coulomb-gas picture 
 \cite{Kada78}
 well describes properties of elementary excitations and also
 singularities of physical quantities
 \cite{Nien84,Blot91,Blot93}. 
 Meanwhile, continuum field theories for systems nearby the criticality
 have been well developed on the basis of the conformal symmetry;
 it offers, beyond the Coulomb-gas description, the unified and
 pervasive approaches to investigate the one-dimensional (1D) quantum
 and two-dimensional (2D) classical systems.
 Further, since they also offer us powerful strategies to numerically
 investigate the systems around the Gaussian fixed point, quantitative
 reliability becomes higher than ever before, which contributes to
 clarify long-standing problems in related research fields
 \cite{Otsu04,Otsu05a,Otsu05b,Mats05}.

 In this paper, we treat the triangular antiferromagnetic Ising model
 (TAFIM) with anisotropic next-nearest-neighbor (NNN) ferromagnetic
 coupling
 \cite{Kita88},
 which we call the Kitatani-Oguchi model hereafter. 
 The reduced Hamiltonian including temperature ${\mathcal H}=\beta H$
 is given by
 \begin{equation}
  {\cal H}(K_1,K_2)
   = K_1 \sum_{\langle j,k\rangle} \delta_{\sigma_j,\sigma_k}
   - K_2 \sum_{      [ j,k]'     } \delta_{\sigma_j,\sigma_k}.
   \label{eq_Hamil}
 \end{equation}
 The binary variable $\sigma_j=0,1$ is on the $j$th site of the
 triangular lattice $\Lambda$ which consists of interpenetrating three
 sublattices $\Lambda_l$ $(l=0,1,2)$. 
 The first sum runs over all nearest-neighbor (NN) pairs $\langle
 j,k\rangle$, whereas the second over NNN pairs $[j,k]'$
 {\it in two of the three directions} (see Fig.\ \ref{FIG1}).
 Although the model was introduced due to a technical reason in
 their numerical calculation, unlike in the isotropic NNN F
 coupling case, it provides an opportunity to investigate an important
 effect, i.e., the spatial anisotropy in critical phenomena.

 Since its introduction, there is a belief that the model\
 (\ref{eq_Hamil}) is not strongly anisotropic and belongs to the
 six-state clock (6SC) universality class as the isotropic model does
 (for more severe anisotropies, see
 \cite{Otsu05b,Noh_94}).
 Thus, two types of Berezinskii-Kosterlitz-Thouless (BKT) transition
 lines are expected to separate the critical region from the ordered and
 disordered phases
 \cite{Bere71,Jose77}.
 Actually, Kitatani and Oguchi (KO) performed the transfer-matrix (TM)
 calculations at the ratio $K_2/K_1=0.5$ in order to evaluate the
 $\beta$ function within the Roomany-Wyld approximation
 \cite{Room80}
 and the spin-spin correlation function, and then concluded the
 existence of the critical finite temperature region
 \cite{Kita88}. 
 Subsequently, Miyashita, Kitatani, and Kanada performed large scale
 Monte Carlo (MC) simulation calculations and found the consistent 
 results 
 \cite{Miya91}.
 Using the cluster TM method (an approximation), Pajersk\'{y} and
 \v{S}urda stated that the intermediate phase possessed an
 incommensurate nature 
 \cite{Paje94}.
 Contrary to these, de Queiroz and Domany performed the TM calculations,
 and reported limited evidence for an existence of the BKT phase on the
 basis of the phenomenological renormalization-group (PRG) analysis
 \cite{Quei95}.
 They argued that this limitation was due to the lack of sixfold
 symmetry, and they also provided the qualitative phase diagram with a
 multicritical point.
 Quite recently, Qian and Bl\"ote succeeded in obtaining the phase
 diagram of TAFIM with isotropic NNN F coupling using both TM and MC
 calculations,  whereas, for present model, they stated that an
 application of the conformal mapping became difficult
 \cite{Qian04}.
 Considering all these together, there still exist some unclear points
 due to the nature of the BKT transitions and/or the effects of the
 anisotropy. 

 In this paper, we shall provide the global phase diagram of the
 Kitatani-Oguchi model\ (\ref{eq_Hamil}) on the basis of the TM
 calculation data.  
 Note especially, our result exhibits that the phase diagram possesses the
 same structure as that of the isotropic model
 \cite{Nien84,Qian04}.
 Therefore, this provides evidence to support KO's motivation and
 confirms the present model being in the 6SC universality class,
 independently of the spatial anisotropy.

 The organization of this paper is as follows: 
 In Sec.\ \ref{sec_EFFEC},
 we shall discuss the effective field theory to describe the low-energy
 and long-distance behaviors of the model, where the effects of the NN
 and the anisotropic NNN couplings are explained. 
 Simultaneously, since we shall employ the level-spectroscopy method to
 treat the two types of BKT transitions
 \cite{Nomu95},
 we address some relevant issues necessary for its application.
 In Sec.\ \ref{sec_RESUL},
 we shall give our numerical estimates of the phase boundary lines.
 By introducing the parameter-dependent geometric factor to obtain
 the isotropic description of the model in the 2D Euclidean space and by
 evaluating the factor on the basis of the conformal field theory (CFT),
 we check the criticality of the system and some universal relations
 among the scaling dimensions.
 The last section, Sec.\ \ref{sec_DISCU},
 is devoted to the discussion and the summary of our research. 
 We provide the analysis of the shapes of boundaries based on the
 crossover argument.
 We also compare our data with previous research results, and then give
 some comments.

\section{Theoretical Description and Numerical Calculation Method}
\label{sec_EFFEC}

\subsection{Effective field theory in the sine-Gordon language}

 We shall start by referring to some relating results in the literature.
 In the exactly solved case $K_2=0$, the system shows the 2D Gaussian
 criticality at $K_1=\infty$.
 In the scaling limit of the lattice spacing $a\to0$, but keeping
 ${\bf x}=(x_1,x_2)=(a j_1, a j_2/\zeta)$ finite [$\zeta=2/\sqrt3$ is
 the geometric factor for the triangular lattice], its effective
 description is given by the Lagrangian density
 \begin{equation}
  {\cal L}_0[\phi]=\frac{1}{2\pi K}[\nabla \phi({\bf x})]^2,
   \label{eq_Gauss}
 \end{equation}
 where $K$ $(=\frac12)$ is the Gaussian coupling and the phase field
 satisfies the periodicity condition $\sqrt2\phi+2\pi=\sqrt2\phi$
 \cite{Nien84,Blot93}.
 Reflecting the discreteness of height variables $h_j$ in the triangular
 Ising solid-on-solid model which is obtained from the zero-temperature
 TAFIM
 \cite{Blot82}, 
 there exists the nonlinear potential $\cos 6\sqrt2\phi$.
 However, it is highly irrelevant at the point
 \cite{Nien84,Blot93},
 so here we dropped it (see below).
 Writing $s_j=e^{i\pi\sigma_j}$ and the distance between $j$th and $k$th
 sites $r_{jk}=|{\bf x}-{\bf x}'|$, then the asymptotic behavior of the
 spin-spin correlation function is given as
 $\langle s_j s_k\rangle \simeq A
 \cos\varphi_{jk}/r_{jk}^{1/2}+B/r_{jk}^{9/2}$,
 where
 $\varphi_{jk}=\varphi_j-\varphi_k$
 and a sublattice dependent phase
 $\varphi_j:=\sum_{l=0,1,2}\sum_{i\in\Lambda_l}({2\pi l}/{3})\delta_{i,j}$ 
 ($A$, $B$ constants.)
 \cite{Step70,Nien84}.
 Thus, the scaling dimensions of the uniform and staggered magnetizations
 $s$ and $S$ constructed through a coarse graining of $s_j$ and $e^{\pm
 i\varphi_j}s_j$ with respect to each elementary triangle are given by
 $x_s=\frac94$ and $x_S=\frac14$, respectively.

 Next, we move on to the anisotropic case $K_2\ne0$. 
 As asserted by KO, there is a plausible reason to expect that
 anisotropic NNN F coupling plays the same role as the isotropic one
 \cite{Kita88},
 i.e., it only reduces the Gaussian coupling down to a certain value
 \cite{Nien84}.
 However, there is no reason to expect that $\zeta=2/{\sqrt3}$ is
 appropriate to obtain a rotationally invariant description of the model
 in the 2D Euclidean space.
 This is because $\zeta=2/{\sqrt3}$ can express the geometric structure
 of the lattice $\Lambda$, but the anisotropic coupling requires a
 further rescaling factor, which must be incorporated
 multiplicatively into the geometric factor.
 Therefore, we suppose that the effective value of the geometric factor
 $\zeta$ is not known {\it a priori} and depending on the coupling values.
 In some systems including those in somewhat different situation, the
 anisotropy effect has been discussed
 \cite{Noh_94,Vill81,Kim_87},
 and further it is also important in the 1D quantum systems described by
 the Tomonaga-Luttinger-liquid Hamiltonian
 \cite{Tomo50}
 \begin{eqnarray}
  H_{\rm TLL}(v,K)
   =
   \int dx
   \frac{v}{2\pi}
   \left[
    K     \left(\partial_x \Theta\right)^2+
    K^{-1}\left(\partial_x   \Phi\right)^2 
   \right],
   \label{eq_FB}
 \end{eqnarray}
 where
 $\left[\Phi(x),\partial_{x'}\Theta(x')/\pi\right]=i\delta(x-x')$, and 
 $v$ is the velocity of an elementary excitation. 
 While, in the imaginary-time formalism, the $(1+1)$ space-time is
 treated as 2D Euclidean, ${\bf x}=(x,v\tau)$, so as to extract the
 Lagrangian density\ (\ref{eq_Gauss}), $v$ as well as $K$ depends on the
 lattice-model parameters.
 Returning to our case, as we will see below, this dependence indeed
 requires an extra procedure to estimate $\zeta$ in numerical
 calculations, but this changes nothing of the theoretical description,
 {\it provided $\zeta$ is chosen properly}.
 Therefore, we advance our argument. 

 For large enough $K_2$, the Gaussian criticality is lifted and the
 sixfold-degenerate ordered state with the $\sqrt3\times\!\sqrt3$
 sublattice structure is stabilized (see Fig.\ \ref{FIG1}).
 On the other hand, for small $K_2$, the thermal scaling field
 $u=e^{-K_1}$ introduces the relevant energy-density perturbation (its
 bosonized expression is $\epsilon=\sqrt2\cos\sqrt2\theta$ in terms of
 $\theta$ being dual to $\phi$) and brings about the disordered state
 \cite{Nien84}.
 In summary, these are described by the dual sine-Gordon model whose
 Lagrangian density is given by
 \begin{equation}
  {\mathcal L}
   =
  {\mathcal L}_0
   +\frac{1}{2\pi\alpha^2}
   \left(y_\phi\cos6\sqrt2\phi+y_\theta\cos\sqrt2\theta\right),
   \label{eq_L}
 \end{equation}
 where $y_\phi<0$ (see below) and $y_\theta\propto u$ are the coupling
 constants
 \cite{Wieg78}.
 $\alpha$ is a short-distance cutoff.
 Since the nonlinear terms in Eq.\ (\ref{eq_L}) are both irrelevant in
 the region $\frac14\ge K\ge\frac19$, the present model\
 (\ref{eq_Hamil}) may possess two BKT-transition lines
 \cite{Jose77}.
 We define $v=e^{K_2}$ and denote the upper- and lower-temperature
 boundaries as $v_{\rm U}(u)$ and $v_{\rm L}(u)$
 [note $v_{\rm U}(u)<v_{\rm L}(u)$]
 for convenience, then their precise determinations are our main goal.

 Quite recently, Matsuo and Nomura investigated the classical 2D 6SC
 model
 \cite{Mats05}. 
 Especially, mapping it to the 1D quantum 6SC model with the explicit
 duality relation, they succeeded in determining two BKT transition
 points, where the crossings of levels observed in the systems with
 periodic and twisted boundary conditions were used. 
 Although their criteria may also be efficient to our case, an
 implementation of the twisted boundary condition is unclear at
 present.
 On the other hand, we have also treated the BKT transitions observed in
 the AF three-state Potts model with the NNN F coupling by employing
 rather naive criteria
 \cite{Otsu05a}.
 Therefore, we shall employ the same approach in the following
 (see also \cite{Otsu05b}).

\subsection{Finite-size estimates of BKT-transition points in upper and
  lower temperatures}\label{subsec_BKT1}
 
 First, let the system around $v_{\rm U}(u)$ be considered where
 $\cos6\sqrt2\phi$ is irrelevant.
 Then, it is well described by 
 \begin{equation}
 {\mathcal L}_1\simeq
  {\mathcal L}_0+
  \frac{y_\theta}{2\pi\alpha^2}\cos\sqrt2\theta~~~
  \left(K\simeq\frac14\right).
 \end{equation}
 With respect to the determination of the BKT-transition points, 
 one of the authors (K.N.) pointed out the importance of marginal
 operators
 \cite{Nomu95}:
 Especially, in this case,
 ${\mathcal M}=(1/K)\left(\nabla \phi\right)^2$
 and
 $\epsilon=\sqrt2\cos\sqrt2\theta$
 hybridize along the RG flow and result in two orthogonalized operators,
 i.e., 
 the ``${\mathcal M}$-like'' and the ``$\cos$-like'' operators. 
 We denote the former and latter as $O_0$ and $O_1$, and
 define the system on $\Lambda$ with $M$ $(\to\infty)$ rows of $L$ (a
 multiple of 3) sites wrapped on a cylinder, say
 $\Lambda(L\times\!\infty)$.
 Then, according to the conformal perturbation theory,
 their renormalized scaling dimensions near the multicritical point
 $(y_0,y_1)=(1/2K-2,y_\theta)=(0,0)$
 are given as
 $x_0(l)\simeq 2-y_0\left(1+\frac43 t\right)$
 and
 $x_1(l)\simeq 2+y_0\left(2+\frac43 t\right)$, respectively, 
 where
 $l=\ln L$
 and a small deviation from the BKT-transition point
 $t=y_1/y_0-1$.
 On the BKT line, $y_0=y_1\simeq 1/l$.
 On the other hand, another important operator is the uniform
 magnetization, $s=\sqrt2\cos3\sqrt2\phi$ in its bosonized form,
 whose dimension is 
 $x_s(l)\simeq\frac9{16}\left(2-y_0\right)$
 in the same region. 
 Consequently, the level-crossing condition,
 \begin{equation} 
  x_0(l)=\frac{16}{9}x_s(l),
   \label{eq_LC1}
 \end{equation}
 provides a finite-size estimate of $v_{\rm U}(u)$, $v_{\rm U}(u,L)$.

 Next, we consider a region near $v_{\rm L}(u)$ where $\cos\sqrt2\theta$
 is irrelevant.
 Then, the effective Lagrangian density is
 \begin{equation}
  {\mathcal L}_2\simeq
   {\mathcal L}_0+
   \frac{y_\phi}{2\pi\alpha^2}
   \cos6\sqrt2\phi~~~
   \left(K\simeq\frac19\right).
 \end{equation}
 In this case,
 ${\mathcal M}$
 and 
 $\sqrt2\cos6\sqrt2\phi$
 hybridize and result in the ${\mathcal M}$-like and $\cos$-like
 operators,
 $O_2$ and $O_3$,
 respectively. 
 By redefining the coupling constants $(y_0,y_1)=(18K-2,-y_\phi)$, we obtain
 the scaling dimensions
 $x_2(l)\simeq 2-y_0\left(1+\frac43 t\right)$
 and
 $x_3(l)\simeq 2+y_0\left(2+\frac43 t\right)$
 near the transition point. 
 Since the uniform magnetization has the dimension
 $x_s(l)\simeq \frac14\left[2-y_0\left(1+2 t\right)\right]$,
 the level-crossing condition for $v_{\rm L}(u,L)$ may be given by
 \begin{equation} 
  x_2(l)=4x_s(l).
   \label{eq_LC2}
 \end{equation}

 Here, let the following be observed: Since we have supposed $y_\phi<0$,
 the sixfold-degenerate ordered states correspond to the six locking
 points of the phase variable, i.e.,
 $\langle\sqrt2\phi\rangle\simeq \pi q/3, (q=0,1,\cdots,5)$. 
 According to the bosonized expressions, the averages of the uniform and
 staggered magnetizations
 [$S={\rm exp}(\pm i\sqrt2\phi)$]
 take values
 $\langle s\rangle\propto e^{i\pi q}$
 and
 $\langle S\rangle\propto e^{\pm i\pi q/3}$
 in the ordered states. 
 These are consistent with the relationship between the phase values and
 the spin configurations given in Refs.\
 \cite{Nien84,Land83}.

\subsection{Discrete symmetry properties of excitations}\label{subsec_Symmet} 

 Now, let us consider the system on $\Lambda(L\times\!\infty)$, and
 define the transfer matrix, ${\bf T}(L)$, connecting the NNN
 rows in the vertical direction of Fig.\ \ref{FIG1}.
 We denote its eigenvalues as
 $\lambda_p(L)$
 or their logarithms as
 $E_p(L)=-\frac12\ln|\lambda_p(L)|$
 ($p$ specifies a level). 
 Then, the conformal invariance provides direct expressions of the
 central charge $c$,
 the scaling dimension $x_p$ and
 the conformal spin $s_p$
 in the critical systems as follows
 \cite{Blot86,Card84}
 \begin{eqnarray}
  E_{\rm g}(L)\simeq Lf-\frac{\pi}{6L\zeta}c,~~~~~~~~~~
  \label{eq_c}\\
  \Delta E_p(L)\simeq \frac{2\pi}{L\zeta}x_p,~~~
  \Delta k_p(L)\simeq \frac{2\pi}{L}s_p.
  \label{eq_xs}
 \end{eqnarray} 
 Here,
 $E_{\rm g}(L)$,
 $\Delta E_p(L)$ $[=E_p(L)-E_{\rm g}(L)]$, and
 $\Delta k_p(L)$
 correspond to the ground-state energy, an excitation gap, and a
 momentum, respectively.
 $f$ is free energy per site.
 Furthermore, the following formulas are available for the Gaussian
 system:
 \begin{eqnarray}
 x_p&&\hspace{-3.5mm}=\frac12\left(Kn^2+\frac{m^2}{K}\right)+(N+\bar N),\\
 s_p&&\hspace{-3.5mm}=nm+(N-\bar N),
 \end{eqnarray}
 where
 $(n,m)$ are the electric and the magnetic charges
 \cite{Kada78},
 and
 $(N,\bar N)$ are non-negative integers
 \cite{U1}.

 In the numerical calculations of the above-mentioned excitation levels,
 the symmetry properties such as
 the translation of one lattice spacing
 (a cyclic permutation among sublattices $\Lambda_i$) ${\cal T}$,
 the space inversion ${\cal P}$,
 and the spin reversal ${\cal S}$
 are quite important. 
 This is because these symmetry operations can be also interpreted
 in the field language as
 ${\mathcal T}: \sqrt2\phi\mapsto \sqrt2\phi+2\pi/3$,
 ${\mathcal P}: \sqrt2\phi\mapsto-\sqrt2\phi       $, and
 ${\mathcal S}: \sqrt2\phi\mapsto \sqrt2\phi+\pi   $
 \cite{Nien84,Blot93,Land83}.
 Therefore, the corresponding levels to marginal operators
  $O_0$ and $O_1$
 ($O_2$ and $O_3$)
 can be found in the subspace of the wavenumber $k=0$ and the even
  parity for both ${\mathcal P}$ and ${\mathcal S}$.
 On the other hand, the uniform magnetization $s=\sqrt2\cos3\sqrt2\phi$ is
 $k=0$ for ${\mathcal T}$ and even for ${\mathcal P}$, but it is odd for
 ${\mathcal S}$ as expected.
 We thus calculate the excitation levels $\Delta E_p(L)$ by utilizing
 these symmetry operations and solve the level-crossing conditions
 (\ref{eq_LC1}) and (\ref{eq_LC2}), numerically. 

 Here, the following should be remarked on.
 When using the KT criterion to determine the BKT-transition points
 [e.g., $K=\frac14$ for $v_{\rm U}(u)$], we should estimate the Gaussian
 coupling from an appropriate excitation gap through the former of Eq.\
 (\ref{eq_xs}).
 If, like the present case, $\zeta$ is not known {\it a priori}, this
 requires its estimate in advance of the gap data. 
 On the other hand, since the level-crossing conditions (\ref{eq_LC1})
 and (\ref{eq_LC2}) are homogeneous expressions in terms of the scaling
 dimensions, the corresponding excitation gaps (or scaled ones), instead
 of the scaling dimensions, can be used to estimate the BKT-transition
 points.
 This property is one of the advantages of the level-spectroscopy
 approach in the studies of 1D quantum and anisotropic 2D classical
 systems. 

 \FIGxSECOND
 \FIGxTHIRD

\section{RESULTS}\label{sec_RESUL}

\subsection{Phase boundaries and the self-dual line}\label{subsec_PBL}

 Now we perform the exact-diagonalization calculations of ${\bf T}(L)$
 for systems up to $L=24$ by the use of the Lanczos algorithm.
 In Figs.\ \ref{FIG2}(a) and  \ref{FIG2}(c), we plot examples of the
 $K_2$ dependencies of the scaled gaps
 $X_p(L):=\Delta E_p(L)/(2\pi/L)$
 (or values multiplied by constants for convenience) at $K_1=1$.
 Then, we can find the points at which the above condition
 (\ref{eq_LC1}) or (\ref{eq_LC2}) is satisfied [i.e., $v_{\rm U,L}(u,L)$]. 
 In addition to the logarithmic corrections, there is another type of
 correction stemming from the $x=4$ irrelevant operators
 \cite{Card86},
 we shall thus extrapolate them to the thermodynamic limit according to
 the least-squares fitting of the polynomial in $1/L^2$.

 In Fig.\ \ref{FIG3}, we give our phase diagram in the 2D model
 parameter space $(u,v)=(e^{-K_1},e^{K_2})$.
 The open circles and squares with the solid curves exhibit the lines
 $v_{\rm U}(u)$ and $v_{\rm L}(u)$, respectively, and they separate an
 intermediate region from the disordered phase and from the ordered
 phase ``Ordered (I)'' with sixfold degeneracy.
 The filled circle at $(u,v)=(1,1+\sqrt2)$ denotes the decoupling point
 with three independent Ising criticality. 
 To complete the phase diagram,
 we also calculate the second-order phase transition point $v_{\rm
 I}(u)$ with the Ising criticality in the F region $u>1$, where the
 finite-size estimates by the PRG method are extrapolated to the
 thermodynamic limit according to 
 $v_{\rm I}(u,L)\simeq v_{\rm I}(u)+a/L^3$ 
 (see the triangles with the solid curve)
 \cite{Nigh76,Derr82}.
 The ordered phase ``Ordered (II)'' has the twofold degeneracy, and the
 thick vertical line shows the first-order phase transition boundary
 between ordered phases.

 For a later discussion, here we mention the self-dual line $v_{\rm sd}(u)$
 embedded in the critical region
 \cite{Lech02}.
 Although it is not the phase transition lines, it is expected to be
 good for numerical calculations
 \cite{Mats05}.
 Defining the transformation
 $6\phi\leftrightarrow\theta$,
 then we can find the duality relation of
 the effective model\ (\ref{eq_L}), i.e.,
 $(K,y_\phi,y_\theta) \leftrightarrow (1/36K,y_\theta,y_\phi)$. 
 Thus it becomes invariant at $K=\frac16$ and $y_\phi=y_\theta$.
 Since the self duality provides the degeneracy of excitation levels,
 e.g., $\cos6\sqrt2\phi$ and $\cos\sqrt2\theta$, 
 their crossing provides the finite-size estimates $v_{\rm sd}(u,L)$.
 However, these are higher energy excitation levels, and their stable
 estimations by the Lanczos method are rather difficult.
 Alternatively, we employ the condition
 $x_{\cal M}(l)=8x_s(l)/3$
 [$x_{\cal M}(=2)$ is the dimension of $\cal M$]; 
 Fig.\ \ref{FIG2}(b) exemplifies the level crossing. 
 Extrapolating them to the thermodynamic limit as
 $v_{\rm sd}(u,L)\simeq v_{\rm sd}(u)+a/L^2$,
 we determine $v_{\rm sd}(u)$
 (diamonds with the dash-dotted line in Fig.\ \ref{FIG3}),
 which is between two boundaries $v_{\rm U,L}(u)$, and is
 terminated at the critical decoupling point.

 \FIGxFOURTH

\subsection{Effective geometric factor, central charge, and consistency
  checks}\label{subsec_Effe}

 Although our calculations so far do not need the numerical estimation
 of $\zeta$, it is necessary to check the criticality and the
 universal relations among scaling dimensions.
 For this, following the recent development in the study of 1D quantum
 systems
 \cite{Kita97}, 
 we focus our attention to the level-1 descendant in the conformal tower
 of the identity operator, i.e., $\hat L_{-1}{\bf 1}$ in the CFT
 language.
 Since the corresponding excitation level is specified by
 $(n,m,N,\bar N)=(0,0,1,0)$,
 we can estimate $\zeta$ as 
 \begin{equation}
  \zeta^{-1}=
   \lim_{L\to\infty}\frac{\Delta E_{\hat L_{-1}{\bf 1}}(L,k=2\pi/L)}{2\pi/L}. 
   \label{eq_V}
 \end{equation}
 The finite-size estimates, i.e., the right-hand side ratio, are
 extrapolated to the thermodynamic limit by the least-squares fitting of
 the polynomial in $1/L^2$.
 Then, we obtain the results which are plotted in Fig.\ \ref{FIG4}(a)
 (the upper panel). 
 Filled triangles plot the data along the $u=0$ line, and open ones 
 along $v_{\rm sd}(u)$.
 Here note that since Eq.\ (\ref{eq_V}) is only valid for the system
 with the Gaussian criticality, it does not hold for others including
 the critical decoupling point.
 For the isotropic case $v=1$, the estimation excellently agrees with
 the exact value, and it increases with $v$ as expected.
 On the other hand, for three decoupled square lattices formed by the
 anisotropic NNN coupling, $\zeta^{-1}$ may equal to $\frac32$, so that
 $\zeta$ seemingly jumps at the critical decoupling point.
 This might have relevance with the jump of the central charge, but a more
 detailed analysis is left for future study.

 Now, according to Eq.\ (\ref{eq_c}), we can estimate the central charge
 $c$ from the $L$ dependence of the ground-state energy $E_{\rm g}(L)$;
 the results are plotted in Fig.\ \ref{FIG4}(b) (the lower panel).
 Although $\zeta^{-1}$ increases to nearly twice as large as the isotropic
 value, the central charge keeps $c=1$ within 1.5\% deviations,
 which clearly demonstrates the Gaussian criticality of the model.

\begin{table}[t]
 \caption{
 Examples of the $L$ dependences of the scaling dimensions
 $x_1$ and $x_4$
 and the averages
 $x_{\rm av}$ and $x'_{\rm av}$
 (see the text) on the BKT-transition points $v_{\rm U,L}(u)$ ($K_1=1$).
 We extrapolate the finite-size estimates to $L\to\infty$ using the
 least-squares fitting of the polynomial in $1/L^2$.
 } 
 \begin{tabular}{rcccccc}
  \hline\hline
    $L$~~~  &12&15&18&21&24&$\infty$\\ 
  \tableline
  $x_1(l)       $ &2.48863&2.50955&2.51917&2.52263&2.52333&     \\
  $x_{\rm av}(l)$ &1.87339&1.90402&1.92525&1.94038&1.95161&1.988\\
  \tableline		
  $x_4(l)       $ &0.83533&0.80457&0.78318&0.76714&0.75449&      \\
  $x'_{\rm av}(l)$ &0.49797&0.49447&0.49202&0.49014&0.48862&0.484\\
  \hline\hline
 \end{tabular}
 \label{TAB_I}
\end{table}

 Next, we shall check some relations.
 Since the amplitudes of the logarithmic corrections are given by the
 operator-product-expansion coefficients, some universal relations among
 the scaling dimensions have been discovered:
 For instance
 \cite{Zima87,Nomu95},
 \begin{eqnarray}
  &\frac13[2x_0(l)+x_1(l)]\simeq      2& \mbox{on~} v_{\rm U}(u),
   \label{eq_xav1}\\
  &\frac14[3x_s(l)+x_4(l)]\simeq\frac12& \mbox{on~} v_{\rm L}(u).
   \label{eq_xav2}
 \end{eqnarray}
 Here, $x_4(l)$ is the scaling dimension of
 $\sqrt2\sin3\sqrt2\phi$,
 which is
 $k=0$ for ${\cal T}$
 and
 odd for ${\cal P}$ and ${\cal S}$. 
 In Table\ \ref{TAB_I}, we give, as an example, the dimensions at
 $K_1=1$ estimated using the former of Eq.\ (\ref{eq_xs}).
 Although
  $x_0(l)$ and $x_1(l)$
 [$x_s(l)$ and $x_4(l)$]
 largely deviate from the free-boson value 2 ($\frac12$) due to the
 logarithmic corrections, their main parts cancel each other in
 Eqs.\ (\ref{eq_xav1}) and (\ref{eq_xav2}).
 Therefore, the average $x_{\rm av}$ ($x'_{\rm av}$), the left-hand
 side of Eq.\ (\ref{eq_xav1}) [Eq.\ (\ref{eq_xav2})], takes the value
 close to 2 ($\frac12$).
 These checks can be passed only if the systems are on the
 BKT-transition lines $v_{\rm U,L}(u)$, and the numerically utilized
 levels have the theoretically expected interpretations.
 Therefore, these are helpful to demonstrate the reliability of our
 numerical results.

\subsection{Summary of results}\label{subsec_Summa}
  
 Consequently, we can confirm that the intermediate phase shows the
 Gaussian criticality with $c=1$ which is separated from the
 sixfold-degenerate ordered and the disordered phases by two
 BKT-transition lines.
 Since these are found not to merge into a single curve, the
 intermediate region as well as the self-dual line continues up to the
 critical decoupling point, and thus it can be regarded as a realization
 of the crossover phenomenon from the criticality $c=\frac32$ at the
 point (we shall discuss this issue in Sec.\ \ref{sec_DISCU})
 \cite{Zamo86}.
 In the previous researches, one expected that the critical region
 was terminated at a certain point, and a first-order phase transition
 between the ordered and disordered phases occurred near the critical
 decoupling point
 \cite{Quei95,Miya91,Meka77}.
 But, this possibility is removed.
 Instead, it is clarified that whereas the obtained phase diagram is, of
 course, quantitatively different from that in the isotropic case, its
 structure, the stabilized phases, and the mechanisms of phase
 transitions are identical to those of the isotropic model
 \cite{Nien84,Qian04}.
 Therefore, we conclude that the present model belongs to the 6SC
 universality class, independently of the spatial anisotropy, and thus
 KO's assertion is confirmed.

\section{DISCUSSIONS}\label{sec_DISCU}

 First, we discuss the nature of our phase diagram
 (in the isotropic case Qian and Bl\"ote performed the similar analysis
 \cite{Qian04}).
 The critical decoupling point with $c=\frac32$ becomes unstable against
 relevant competing perturbations and exhibits crossovers to the
 behaviors controlled by the critical fixed points with lower symmetries
 \cite{Zamo86}.
 While one of those perturbations is the energy-density operator of the
 Ising model with the dimension 1, another one may be a product of the
 magnetization operators on two of three sublattices which has the
 dimension $2\times\frac18$
 \cite{Leeu75}.
 Therefore, the crossover exponent
 $\phi=(2-\frac14)/(2-1)=\frac74=1.75$
 can predict the shape of the boundary around the point. 
 In Fig.\ \ref{FIG5}
 the log-log plot of the phase boundary line $v_{\rm I}(u)$ is given by
 the triangles with the least-squares-fitting line to the data for the
 four smallest $|K_1|$ [we define $K^*_2=\ln(1+\sqrt2)$].
 The inverse of the slope estimates the exponent
 $\phi_{\rm F}\simeq 1.72$. 
 We also analyze $v_{\rm sd}(u)$, i.e., the crossover line to the
 Gaussian fixed point with $K=\frac16$ (see the diamonds with the
 fitting line in the same figure).
 Then, the estimated exponent also takes the close value, i.e.,
 $\phi_{\rm AF}\simeq 1.76$.
 Therefore, we can confirm the above crossover argument. 
 On the other hand, the analysis of the BKT-transition lines $v_{\rm
 U,L}(u)$ becomes problematic.
 This may be mainly due to the existence of the corrections stemming
 from the marginal operators, while those are absent in the F case.
 
 \FIGxFIFTH

 Second, based on our results obtained in the above, we shall provide some
 comments on previous work: 
 Miyashita, Kitatani, and Kanada
 performed the MC simulations at $K_2/K_1=0.2$ and 0.5
 \cite{Miya91}. 
 While they used some methods to estimate the transition points, the
 most specific ones are those by the MC-RG method
 \cite{Kiku87}.
 They employed the real-space renormalization for each elementary
 triangle with corner spins $s_1,s_2,s_3$, i.e., 
 $s_1+s_2+s_3-s_1s_2s_3 \mapsto s'$
 \cite{Niem73},
 and then evaluated two BKT-transition points. 
 The comparison between Fig.\ \ref{FIG3} and their MC-RG data exhibits
 that the lower-temperature transition point at $K_2/K_1=0.5$ deviates
 from our phase boundary.
 One may attribute the discrepancy to the statistical errors in MC
 simulations, but there might be a possibility that the anisotropy
 effect could increase an uncontrollability of the real-space RG
 treatment.

 Pajersk\'y and \v{S}urda
 employed the cluster TM method (a combination of the TM and a
 mean-field approximation), and claimed that the intermediate critical
 phase possessed an incommensurate structure
 \cite{Paje94}. 
 They also implied that the lower-temperature transition between an
 incommensurate liquid and the commensurate ordered phase was the
 Pokrovsky-Tarapov type
 \cite{Pokr79}. 
 However, as exhibited, the dual sine-Gordon field theory\ (\ref{eq_L})
 well describes the phase transitions observed in this model, and there
 exists no relevant term to stabilize the incommensurate phase
 \cite{Alle01}.
 Further, since the duality relation is possessed by the effective model
 ($u\ne0$) and it interchanges the transition points in upper and lower
 temperatures, they should be the same type.
 Therefore, their observations may be due to an artifact in their
 mean-field approximation treatment.
 
 de Queiroz and Dommany
 investigated the same model by the TM calculations
 \cite{Quei95}. 
 They provided limited evidence for the existence of the BKT phase, and
 also alternative scenarios including an absence of the BKT phase.
 Meanwhile, based on the PRG calculation data and the scaled gaps, they
 drew the qualitative phase diagram including the multicritical point
 and the direct transition line between ordered and disordered phases. 
 However, it has been pointed out by several authors that the PRG
 calculations fail to estimate the BKT-transition points
 \cite{Inou99}; 
 their results suffer from an inadequacy of the method.
 Further, they tried to check the KT criterion on their phase
 boundaries. 
 They indeed estimated the exponent $\eta$ $(=2x_S)$ by the use of Eq.\
 (\ref{eq_xs}) and the fixed value $\zeta=2/\sqrt3$.
 But, as we have seen, $\zeta$ depends upon model parameters, so the
 scaled gaps cannot give their universal amplitude in the anisotropic
 case.
 Furthermore, their observation that the ratio of scaled gaps gives
 the flattening region of the parameter is understandable from our
 viewpoint that approximate cancellation of the geometric factor
 occurs by taking their ratio.
 Consequently, these exhibit that the nonuniversal quantity $\zeta$ is
 important in the quantitative description using the numerical
 calculations, although it is not theoretically important.

 To summarize,
 we investigated the Kitatani-Oguchi model by the level-spectroscopy
 method in order to clarify spatial anisotropy effects and the global
 phase diagram.
 By taking into account the parameter dependence of the geometric factor
 properly, we analyzed the scaling dimensions of operators around the
 BKT-transition lines.
 Then, we numerically determined the phase diagram, where two types of
 Berezinskii-Kosterlitz-Thouless transition lines separate the  
 intermediate critical phase from the ordered and disordered phases. 
 Further, we evaluated the central charge and performed consistency
 checks among scaling dimensions in order to provide evidence of
 the universality class, and then we confirmed the assertion made by
 Kitatani and Oguchi.
 Some comments and comparisons with previous work were also given
 on the basis of our viewpoint and numerical results. 

 While our approach has its basis on the argument of the
 Tomonaga-Luttinger liquid observed in the 1D quantum systems, it is
 widely applicable to the 2D classical systems with spatial anisotropy.

\section*{Acknowledgments}
 One of the authors (H.O.) thanks
 H. Matsuo,
 M. Nakamura,
 K. Okunishi, 
 M. Fujimoto, 
 and 
 T. Yokoo
 for stimulating discussions. 
 We also appreciate
 H. Kitatani
 for reading our manuscript prior to submission.
 Main computations were performed using the facilities of
 Information Synergy Center, Tohoku University.
 This work was supported by Grants-in-Aid from
 the Japan Society for the Promotion of Science.

\end{document}